\newcommand{\mathm}[1]{{\mathbf{\mathsf #1}}}
\newcommand{\mathv}[1]{{\mathbf{\mathbf #1}}}

\documentstyle[epsfig]{mn}

\title[Multivariate Non--Normality in the WMAP 1st Year Data]{Multivariate Non--Normality in WMAP 1st Year Data}

\author[P. Dineen \& P. Coles]{Patrick Dineen$^{1,2}$\thanks{E-mail: p.dineen@ic.ac.uk} \& Peter Coles$^{2}$\\
$^{1}$Astrophysics Group, Blackett Laboratory, Imperial College, Prince Consort Road, London SW7 2AZ, United Kingdom\\
$^{2}$School of Physics \& Astronomy, University of Nottingham, University Park, Nottingham, NG7 2RD, United Kingdom\\
}

\begin{document}
\maketitle

\begin{abstract}
The extraction of cosmological parameters from microwave background
observations relies on specific assumptions about the statistical
properties of the data, in particular that the $p$-point
distributions of temperature fluctuations are jointly--normal. Using
a battery of statistical tests, we assess the multivariate Gaussian
nature of the Wilkinson Microwave Anisotropy Probe (WMAP) 1st year
data. The statistics we use fall into three classes which test
different aspects of joint--normality: the first set assess the
normality of marginal (one-point) distributions using familiar
univariate methods; the second involves statistics that directly
assess joint--normality; and the third explores the evidence of
non--linearity in the relationship between variates. We applied
these tests to frequency maps, `foreground--cleaned' assembly maps
and all--sky CMB--only maps. The assembly maps are of particular
interest as when combined with the kp2 mask, we recreate the region
used in the computation of the angular power spectrum. Significant
departures from normality were found in all the maps. In particular,
the kurtosis coefficient, D'Agostino's statistic and bivariate
kurtosis calculated from temperature pairs extracted from all the
assembly maps were found to be non--normal at 99\% confidence level.
We found that the results were unaffected by the size of the
Galactic cut and were evident on either hemisphere of the CMB sky.
The latter suggests that the non--Gaussianity is not simply related
to previous claims of north--south asymmetry or localized
abnormalities detected through wavelet techniques.
\end{abstract}


\section{Introduction}
\label{sec:intro}

Our picture of the Universe has evolved remarkably over the past
century; from a view limited to our Galaxy to a cosmos with billions
of similar structures. Today's  era of ``precision cosmology'',
cosmology appears to be concerned with improving estimates of
parameters that describe the standard cosmological model rather than
testing for possible alternatives. Recent observations of the large
scale structure (eg. 2dFGRS; Percival et al. 2001) and the Wilkinson
Microwave Anisotropy Probe (WMAP; Bennett et al. 2003a) observations
of the Cosmic Microwave Background (CMB) appear to confirm our key
ideas on structure formation. However, there is the suggestion that
our confidence may be misplaced. The WMAP observations, for example,
have thrown up a number of unusual features
\cite{cnvw03,e03,ndv03,cn04,cdew04,dc04,ehbg04,lw04,p04,lm05a,mc05}. In
particular, it has been noted that the lowest spherical harmonics
behave in a peculiar way, with an anomalously low quadrupole (see
Efstathiou 2003, 2004) and confusion over the planar nature and
alignment of the quadrupole and octopole \cite{tdh03,dtzh04}.
Multipole vectors have also been used to show that there is
preferred direction in the CMB sky \cite{chs04,lm05b}. Wavelet
techniques have demonstrated that particular regions in the sky
display unusual characteristics (eg. Vielva et al. 2004;  McEwen et
al. 2005). Other statistical analyses give results that are
compatible with Gaussianity (Colley \& Gott 2003; Komatsu et al.
2003; Stannard \& Coles 2005). There is some danger here of a
``publication bias'' in which only positive detections are reported
so and it appears a good time to stand back from the established
paradigm and systematically test the core assumptions intrinsic to
our view of the universe.

Standard cosmological models assume the structures we see today grew from
small initial perturbations through gravitational instability.
The initial perturbations are thought to be seeded during a period of
inflation \cite{g81,as82,l82}.
These primordial perturbations are the result of amplification of zero--point quantum
fluctuations to classical scales during the inflationary period \cite{gp82,h82,s82}.
In most inflationary scenarios the resultant primordial density field is taken to
be a statistically homogeneous and isotropic Gaussian random field
\cite{adler81,bbks86}. Such a field has both physical and mathematical
attractions. However, there are alternative possibilities that lead to a
field with non--Gaussian statistics. Versions of inflation
involving multiple scalar fields or those with non--vacuum initial states
lead to a non--Gaussian spectrum.
Other potential candidates for seeding the primordial field also lead to
non--Gaussian possibilities. Topological defects arise from
symmetries being broken as the early Universe cools. The topological defect scenario includes
models based on cosmic strings, textures and monopoles
\cite{k76,vilenkin94}. The development of the defects involves nonlinear physics
which leads to a non--Gaussian pattern of fluctuations.

The question of non--Gaussianity in the CMB is intertwined with the
same question concerning the initial density perturbations. The
inhomogeneity in the matter distribution is translated to the
background radiation photons through Thomson scattering. After last
scattering the majority of these photons arrive at our observatories
unperturbed. If the primordial matter distribution is Gaussian then
so too is the primary temperature pattern. Thus, the CMB
anisotropies provides us with a clean tool to discriminate between
initial field models described in the previous section. However,
before we can test this assumption we need to ensure we are looking
at the last--scattering surface. Most secondary anisotropies are
nonlinear in nature and hence lead to non--Gaussian signals. Lensing
and the Sunyaev--Zel'dovich effect should both be able to be
isolated through their non--Gaussian effects. Moreover, both
Galactic foregrounds and experimental systematic errors will leave
non--Gaussian imprints in the CMB measurements. Thus, the
development of non--Gaussian statistics that can isolate these
effects is imperative. The demand for such tracers will only
intensify: increased sensitivity and polarization measurements will
require better control of secondary effects and foregrounds.

One of the primary goals of the WMAP mission is to make precise
estimates of the cosmological parameters. These parameters are
extracted by comparing the measured angular power spectrum to
cosmological model predictions. The statistical information of the
(processed) temperature field is assumed to be entirely encoded in
the power spectrum, which is true if the field is a multivariate
Gaussian. It is the aim of this paper to investigate this
assumption: if it does not hold we may have to question the
inferences obtained from the spectrum and address the causes of this
non--normality whether cosmological, Galactic or systematic.

In this paper we approach this issue from a general statistical
standpoint. Rather than looking for specific non-Gaussian
alternatives we simply explore the extent to which the assumption of
multivariate Gaussianity is supported by the data using the most
general approaches available. The layout of the paper is as follows.
In the next section we clarify the definition of a multivariate
Gaussian random field. In Section \ref{sec:analysis}, we outline
techniques we shall use to probe the normality of current CMB data
sets from WMAP. In Section \ref{sec:data}, we discuss the nature of
the CMB data sets that the techniques are applied to. The details of
the implementation of the method are drawn in
\ref{sec:implementation}. In Section \ref{sec:results}, we present
the results of our analysis. The conclusions are discussed in
Section \ref{sec:conclusion}.

\section{Gaussian Random Fields}

Gaussian random fields arise in many situations that require the
modelling of stochastic spatial fluctuations. There are two reasons
for their widespread use. One is that the Central Limit Theorem
requires that, under very weak assumptions, that linear additive
processes tend to possess Gaussian statistics. To put this another
way, the normal distribution is the least ``special'' way of
modelling statistical fluctuations. The other reason is that
Gaussian processes are fully specified in a mathematical sense. Very
few non-Gaussian processes are as tractable analytically. The
assumption of multivariate Gaussianity is consequently the simplest
starting point for most statistical analyses.

If we signify the fluctuations in a Gaussian random field by $\delta(\mathv{r})=\delta$,
then the probability distribution function of $\delta$ at individual spatial positions
is given by
\begin{equation}
{\cal P}(\delta) = \frac{1}{(2\pi \sigma^2)^{1/2}}\exp\left( -\frac{\delta^2}{2\sigma^2} \right)
\end{equation}
where $\sigma^2$ is the variance and $\delta$ is assumed to have a
mean of zero. In fact, the formal definition of a Gaussian random
field requires all the finite-dimensional $p$--variate joint
distribution of a set of $\delta(\mathv{r}_i)=\delta_i$ to have the
form of a multivariate Gaussian distribution:
\begin{equation}
\label{eqn:multigauss} {\cal
P}(\mathv{\Delta})=\frac{|\mathm{\Sigma}^{-1}|^{1/2}}{(2\pi)^{p/2}}\exp\left(
-\frac{1}{2}\mathv{\Delta}'\cdot\mathm{\Sigma}^{-1}\cdot\mathv{\Delta}\right),
\end{equation}
where $\mathv{\Delta}=(\delta_1,\delta_2,\dots,\delta_p)'$ and $\Sigma$ is the covariance matrix \cite{k00}.
Therefore, knowledge of the mean and variance of each variate,
and all covariances between pairs of variates, fully specifies the field.

So how does this formalism manifest itself when we are dealing with the CMB? The temperature fluctuations $\Delta T(\theta, \phi)$ in the CMB at any point in the celestial sphere should be drawn from a multivariate Gaussian. That is to say, ${\cal P}(\Delta T_1, \Delta T_2, \ldots, \Delta T_p)$ should have the same form of Equation \ref{eqn:multigauss}. The CMB fluctuations can also be expressed in spherical harmonics as
\begin{equation}
\Delta T(\theta, \phi)=\sum_{\ell=1}^{\infty }\sum_{m=-\ell}^{m=+\ell}a_{\ell m}Y_{\ell m}(\theta ,\phi ),
\end{equation}
where the $a_{\ell m}$ are complex and can be written
\begin{equation}
a_{\ell m}=|a_{\ell m}|\exp[i\varphi_{\ell m}].
\end{equation}
If the temperature field is a multivariate Gaussian then the real and imaginary parts of the $a_{\ell m}$ should be mutually independent and Gaussian distributed \cite{bbks86}. Furthermore, the phases $\varphi_{\ell m}$ should be random. In our analyses we look for non--Gaussianity in the WMAP data in both real and harmonic spaces. Both spaces allows us to probe a range of length scales. Harmonic space has the advantage of being more condensed. Whereas, the advantage of studying in real space is that we can easily navigate contaminated regions or focus on a specific area in the CMB sky. The exact nature of the data used is fully described in Section \ref{sec:data}.

\section{Multivariate analysis}
\label{sec:analysis}

In this Section, we outline the procedures used later to examine
whether the WMAP data is strictly multivariate Gaussian. Generally,
when looking for signs of non-normality, it helps to have an idea of
the form it should take. The wide variety of possible sources of
non--Gaussianity means that there is no unique form of alternative
distribution to seek. Fortunately, this is quite a common problem in
multivariate analyses, where real data does not adhere to any
specific alternative model. This is unsurprising as there are very
alternative models for which the entire hierarchy of multivariate
distributions is fully specified. For this reason statisticians tend
to apply not just one test statistic to the data, but a battery of
complementaty procedures. The assorted procedures will have
differing sensitivities to the shape of the distribution. There is
also a need to augment these tests with analyses of subsets of the
data. Testing the form Equation \ref{eqn:multigauss} in all its generality is clearly
impossible as it requires an infinite number of tests. In practice,
various simplified approaches tend to be implemented: in particular,
the one-point marginal distributions of the variates are often
studied. Marginal normality does not imply joint normality, although
the lack of multivariate normality is often reflected in the
marginal distributions. A further advantage of examining the
marginal distributions is that they are computationally less intense
and more intuitive (i.e. easier to interpret) and thus more
instructive.

The procedures we apply to the data are described in three
subsections. In subsection \ref{sec:univariate}, we outline
univariate techniques for assessing marginal normality. In
subsection \ref{sec:multivariate}, multivariate techniques for
evaluating joint normality are sketched out. Lastly, we illustrate a
procedure that evaluates the degree to which the regression of each
variate on all others is linear in subsection \ref{sec:linear}.
Throughout these subsections, we shall denote the members of the
$i$th variate, $\mathv{x}_i$, by $x_{ij}$ where $j=1,\dots,n$. In
subsection \ref{sec:multivariate}, we shall use the notation
$\mathv{x}_j=(x_{1j}, x_{2j}, \dots, x_{pj})'$ and similarly
$\mathv{x}_k=(x_{1k}, x_{2k}, \dots, x_{pk})'$. We wish to emphasize
the distinction between $\mathv{x}_j$ and $\mathv{x}_i=(x_{i1},
x_{i2}, \dots, x_{in})'$.

\subsection{Evaluating marginal normality}
\label{sec:univariate}

The evaluation of marginal normality of the data is based on well--known
tests of univariate normality. The marginal distributions we study
correspond to the distribution for each individual variate.
This is simply the distribution of members of the specified variate,
ignoring all other members from the data--set. For example, in the
bivariate case, the marginal probability distribution of the
variate $\mathv{x}_1$ is given by
\begin{equation}
\label{eqn:marginal}
{\cal P}(\mathv{x}_1)=\int_{-\infty}^{\infty}{\cal P}(\mathv{x}_1,\mathv{x}_2) d\mathv{x}_2.
\end{equation}
The parallel expression for the marginal distributions corresponding
to larger values of the dimensionality $p$ can easily be developed.
We outline four techniques that probe univariate normality of ${\cal
P}(\mathv{x}_i)$: the skewness and kurtosis coefficients;
D'Agostino's omnibus test; and a shifted--power transformation test.
In this subsection, we shall suppress the $i$ indices when referring
to the data--members such that $x_j$ will refer to the individual
members of $\mathv{x}_i$.

The classic tests of normality is by means of evaluating the sample
skewness $\sqrt{b_1}$ and kurtosis $b_2$ coefficients. If we let the
sample mean and the sample variance be $\bar{x}$ and $S^2$
respectively, and define
\begin{equation}
m_r=\frac{1}{n}\sum_{j=1}^{n}\left( x_j-\bar{x} \right)^r,
\end{equation}
to be the $r^{\mathrm th}$ moment about the mean. Then the skewness is given by
\begin{equation}
\sqrt b_1=\frac{m_3}{S^3},
\end{equation}
and the kurtosis by
\begin{equation}
b_2=\frac{m_4}{S^4}.
\end{equation}
The expected value of the skewness \cite{m80} is
\begin{equation}
E(\sqrt b_1)=0,
\end{equation}
and the square root of the variance of this quantity is
\begin{equation}
\sigma(\sqrt b_1)=\sqrt{\frac{6}{n}} \cdot \left( 1- \frac{3}{n} + \frac{6}{n^2} - \frac{15}{n^3} + \cdots \right).
\end{equation}
Here, and throughout the rest of the paper, we use the notation
$E(Q)$ for the expected value of the statistic $Q$ and $\sigma(Q)$
to the signify the square root of the variance about this value.
These values assume $Q$ is applied to a Gaussian data--set. The same
quantities for the kurtosis \cite{m80} are
\begin{equation}
E(b_2)=\frac{3(n-1)}{n+1},
\end{equation}
and
\begin{equation}
\sigma(b_2)=\sqrt{\frac{24}{n}} \cdot \left( 1- \frac{15}{2n} + \frac{271}{8n^2} - \frac{2319}{16n^3} + \cdots \right).
\end{equation}

So we have the expected value and variance of the kurtosis and
skewness. But how useful are these quantities? Are we justified in
using Gaussian statistics to measure non--normality? As the value of
$n$ increases the distribution of $E(\sqrt b_1)$ soon reverts to a
normal distribution. However, the distribution of $E(b_2)$ is very
skewed for $n$=100 and is hardly normal for $n$=1000 \cite{m80}.
Clearly large values of $n$ are required to be confident of the any
analysis using the kurtosis. This last point is addressed further in
Section \ref{sec:data} when our sample sizes are discussed.

There are a number of general tests that can be used to monitor the
shape of the distribution of a variate. Some of these try to combine
the skewness and kurtosis coefficients into an omnibus test. Other
statistics aim to probe different features of the shape of the
parent distribution. Using these statistics tend to be problematic
as they usually require tabulated coefficients that are cumbersome
and not available for large $n$. For this reason D'Agostino (1971)
developed an omnibus test of normality for large sample sizes based
on order statistics. The test statistic is defined as
\begin{equation}\label{eqn:dagostino}
D=\frac{1}{n^2S}\sum_{j=1}^{n}\left(j-\frac{1}{2}(n+1)\right)x_{(j)},
\end{equation}
where $x_{(1)}$, $x_{(2)}$, $\dots$, $x_{(n)}$ are the $n$ order statistics of the sample such that $x_{(1)} \leq x_{(2)} \leq \dots \leq x_{(n)}$. The expectation is
\begin{equation}
E(D)\simeq\frac{1}{2\sqrt\pi},
\end{equation}
and the variance asymptotically is given by
\begin{equation}
\sigma(D)=\frac{0.02998598}{\sqrt n}.
\end{equation}

So far, we have looked at descriptive measures of the normality of the
marginal distribution. There are also tests based on transforming the data.
Box \& Cox (1964) proposed using a shifted--power transformation to assess normality
\begin{equation}
x_j\rightarrow x_j^{(\xi,\lambda)}=\left\{ \begin{array}{ll}
        \left( (x_j+ \xi)^{\lambda} -1 \right)/\lambda & \mbox{$\lambda \neq 0$, $x_j>-\xi$}\\
        \log(x_j+\xi) & \mbox{$\lambda= 0$, $x_j>-\xi$}\end{array} \right.
\end{equation}
The transformation aims to improve the normality of the variate $x$;
$\lambda$ can be estimated by maximum likelihood and the null
hypothesis, $H_0\colon\lambda=1$, tested by a likelihood ratio test.
The shift parameter $\xi$ is included in the transformation because
it appears to respond to the heavy--tailedness of the data, whereas,
$\lambda$ appears sensitive to the skewness \cite{g97}.
Specifically, it can be shown that the log--likelihood function is
given by
\begin{equation}
\label{eqn:likelihood}
{\cal L_{\mathrm{max}}}(\xi,\lambda)=-\frac{n}{2} \ln \hat{\sigma}^2 + (\lambda-1)\sum_{j=1}^{n}\ln(x_j+\xi)
\end{equation}
where $\hat{\sigma}^2$ is the maximum likelihood estimate of the transformed
distribution that is presumed to be normal such that
\begin{equation}
\hat{\sigma}^2=\frac{1}{n}\sum_{j=1}^{n}\left( x_j^{(\xi,\lambda)} - \bar{x_j}^{(\xi,\lambda)}\right)^2
\end{equation}
where
\begin{equation}
\bar{x_j}^{(\xi,\lambda)}=\frac{1}{n}\sum_{j=1}^{n}x_j^{(\xi,\lambda)}
\end{equation}
Equation \ref{eqn:likelihood} is maximised to obtain the maximum
likelihood estimates $\hat{\xi}$ and $\hat{\lambda}$. Finally, a
significance test can be constructed by comparing the value of
$2\!\left({\cal L_{\mathrm{max}}}(\hat{\xi},\hat{\lambda})-{\cal
L_{\mathrm{max}}}(\hat{\xi},1)\right)$ to a $\chi^2$ distribution
with one degree of freedom. Note that ${\cal
L_{\mathrm{max}}}(\xi,1)$ is independent of $\xi$ so this is a free
parameter in the test.

\subsection{Evaluating joint normality}
\label{sec:multivariate} As we have already remarked, deviations
from joint normality should be detectable through methods directed
toward testing the marginal normality of each variate. However,
there is a need to explicitly test the multivariate nature of the
data. Therefore, we look at three techniques that fulfil this
requirement. These techniques are multivariate generalizations of
univariate tests outlined in Section \ref{sec:univariate}.

The first two techniques to evaluate joint normality are multivariate
measures of skewness and kurtosis. These methods were developed by
Mardia (1970) and make use of Mahalanobis distance of $\mathv{x}_j$ and
Mahalanobis angle between $\mathv{x}_j-\mathv{\bar{x}}$ and $\mathv{x}_k-\mathv{\bar{x}}$.
The Mahalanobis distance is defined as
\begin{equation}
r_j^2=(\mathv{x}_j-\mathv{\bar{x}})'\mathm{S}^{-1}(\mathv{x}_j-\mathv{\bar{x}}),
\end{equation}
where ${\mathm{S}}$ is the sample covariance matrix. It is often used to assess the
similarity of two or more data sets. The Mahalanobis angle is given by
\begin{equation}
r_{jk}=(\mathv{x}_j-\mathv{\bar{x}})'\mathm{S}^{-1}(\mathv{x}_k-\mathv{\bar{x}}).
\end{equation}
The multivariate skewness is related to the Mahalanobis angle and as such reflects
the orientation of the data. It is defined as
\begin{equation}
b_{1p}=\frac{1}{n^2}\sum_{j=1}^{n}\sum_{k=1}^{n}r_{jk}^3,
\end{equation}
where $nb_{1p}/6$ is asymptotically distributed as a $\chi^2$ with $p(p+1)(p+2)/6$ degrees of freedom.
In the bivariate case, this is a $\chi_4^2$ distribution.
The multivariate kurtosis is defined as the mean of the square of the Mahalanobis distance
\begin{equation}
b_{2p}=\frac{1}{n}\sum_{j=1}^{n}r_j^4.
\end{equation}
It is expected to be normally distributed with mean $p(p+2)$ and variance $8p(p+2)/n$.
Therefore, we have $E(b_{22})=8$ and $\sigma(b_{22})=\sqrt{64/n}$.

Our last multivariate technique is an extension of the transformation test.
We simply look at just the power transformation of each variable separately
ignoring the shift parameter $\xi$. That is to say
\begin{equation}
\mathv{x}_i\rightarrow \mathv{x}_i^{(\mathv{\lambda})}=\left\{ \begin{array}{ll}
        \left( x_{ij}^{\lambda_i} -1 \right)/\lambda_i & \mbox{$\lambda_i \neq 0$}\\
        \log(x_{ij}) & \mbox{$\lambda_i= 0$}\end{array} \right.
\end{equation}
The maximised log--likelihood function is given by
\begin{equation}
\label{eqn:mlikelihood}
{\cal L_{\mathrm{max}}}(\mathv{\lambda})=-\frac{n}{2} \ln| \mathm{\hat{\Sigma}}| + \sum_{i=1}^{p}\left[ (\lambda_i -1) \sum_{j=1}^{n} \ln x_{ij} \right],
\end{equation}
where $\mathm{\hat{\Sigma}}$ is maximum likelihood estimate of the
covariance matrix of the transformed data \cite{g97}. $\mathm{\hat{\Sigma}}$ is
computed in analogous fashion to $\hat{\sigma}^2$ in Equation \ref{eqn:likelihood}.
The transformation with $\mathv{\lambda}=(\lambda_1,\dots,\lambda_p)'=\mathv{1}$
is the only transformation consistent with normality. Therefore, as with the
univariate case, we can compare the value of
$2\!\left({\cal L_{\mathrm{max}}}(\mathv{\hat{\lambda}})-{\cal L_{\mathrm{max}}}(\mathv{1})\right)$
to a $\chi_p^2$ distribution.

\subsection{Evaluating linearity}
\label{sec:linear} There is a further aspect of multivariate
normality that we can examine in our data set: all variates, if not
independent of each other, should be linearly related. Here we are
not specifically testing the normality of the variates, as it is
easy to imagine variates that are extremely non--normal yet linearly
related (eg. two identical sawtooth distributions). Importantly
though, any non--linearity does inform us about the usefulness of
the covariance matrix. Whereas, marginal non-normality may allow you
to still extract useful information from the covariance matrix.
Non-linearity results in the more serious consequence of the
covariance matrix being a poor indicator, even qualitatively, of the
association of variates \cite{bs78}. Our prime motivation for
assessing the multivariate Gaussianity of the WMAP data is due to
statistical nature of the data underpinning the extraction of
cosmological parameters. It is therefore essential that the
covariance matrix fully specifies all the information from the data
set. Clearly, testing linearity is not only complementary to the
previous tests we have outlined, but also can be thought of as a
vital step in assessing the Gaussianity of CMB data.

Investigating the linearity of regression of variates as a means of
scrutinizing multivariate normality was first proposed by Cox \&
Small (1978). We shall outline a related method for obtaining and
evaluating regression coefficients described in Montgomery (1997).
To begin with, let us look at the general concepts behind multiple
linear regression. If we have $n$ measurements of two variables $y$
and $z$, then, the two variables are related linearly when the
following model fully describes their relationship
\begin{equation}
y_j=A_0 + A_1 z_j + \epsilon_j,
\end{equation}
where $\epsilon_j$ is the error term in the model and (as usual) the
subscript $j$ runs from 1 to $n$. The parameters $A_{\alpha}$ are
referred to as regression coefficients. We can consider building other terms into our model
\begin{equation}\label{eqn:nonlinear}
y_j=A_0 + A_1 z_{j1} + A_2 z_{j2} + \cdots + A_q z_{jq} +
\epsilon_j,
\end{equation}
where $z_{j\alpha}$ can be $z_j^2$, $z_jy_j$, $z_j^{1/3}$ and so on.
Linearity dictates that the regression coefficients $A_{\alpha}$
should be zero for all non--linear terms. So how do we estimate
$A_{\alpha}$? A typical technique is to minimize the error term
$\epsilon_j$. If we rewrite Equation (\ref{eqn:nonlinear}) in matrix
notation, we have
\begin{equation}
\mathv{y}=\mathm{Z}\mathv{A} + \mathv{\epsilon}
\end{equation}
where $\mathv{A}=(A_0,\cdots,A_q)'$ and $\mathm{Z}$ is a
$n\!\times\! (q+1)$ matrix with the first column comprised of ones
and the subsequent columns corresponding to $k$ regressor variables
in the model. The error term can be minimized using the method of
least--squares. The least--squares function is given by
\begin{equation}\label{eqn:leastsquare}
L=\sum_{j=1}^{n}\epsilon_j=\mathv{\epsilon}'\mathv{\epsilon}=(\mathv{y}-\mathm{Z}\mathv{A})'(\mathv{y}-\mathm{Z}\mathv{A}).
\end{equation}
Differentiating Equation \ref{eqn:leastsquare} with respect to
$\mathv{A}$ and equating this to zero, minimizes this function and
leads us to
\begin{equation}
\mathv{\hat{A}}=\left( \mathm{Z}'\mathm{Z} \right)^{-1}\mathm{Z}'\mathv{y},
\end{equation}
which is our best estimate of $\mathv{A}$.

Once we have obtained a set of regression coefficients, we wish to
test the significance of each coefficient. If the variables are
linearly related then the coefficients for the additional
coefficients to our model should be zero. Given an individual
coefficient $A_{\alpha}$, we therefore want to check the hypothesis
$H_0\colon A_{\alpha}=0$ against $H_1\colon A_{\alpha}\neq0$. If
$H_0$ is not rejected for one of the additional coefficients, then
the term needs to be added to model the behaviour of the variables.
Montgomery (1997) outlines a statistic $t_0$ that can be used to
test this hypothesis. The error in our estimate of $\mathv{\hat{A}}$
is given by
\begin{equation}
\hat{\sigma}^2=\frac{1}{n-q}\cdot\left( \mathv{y}'\mathv{y}-\mathv{\hat{A}}'\mathm{Z}'\mathv{y}  \right).
\end{equation}
The test statistic for $H_0$ is
\begin{equation}
\label{eqn:ttest}
t_0=\frac{\hat{A}_{\alpha}}{\sqrt{\hat{\sigma}^2C_{\alpha\alpha}}}
\end{equation}
where $C_{\alpha\alpha}$ is the diagonal element of $(\mathm{Z}'\mathm{Z})^{-1}$ corresponding
to $A_{\alpha}$. The statistic $t_0$ should match a $t$--distribution with $(n-q-1)$ degrees of freedom.

\section{Data}
\label{sec:data}

The WMAP instrument comprises $10$ differencing assemblies
(consisting of two radiometers each) measuring over 5 frequencies
($\sim\! 23$, $33$, $41$, $61$ and $94$ GHz). The two lowest frequency
bands ($K$ and $Ka$) are primarily Galactic foreground monitors, while
the highest three ($Q$, $V$ and $W$) are primarily cosmological bands
\cite{hsvh03}.

In our study, we look at a total of 15 maps constructed from the
WMAP data. All the maps were obtained from the NASA's
LAMBDA\footnote{http://lambda.gsfc.nasa.gov} data archive. The maps
are in HEALPix\footnote{http://healpix.jpl.nasa.gov} format with a
resolution parameter of $N_{\mathrm{side}}$=512. The data consists
of: 5 frequency maps; 8 'foreground--cleaned' assembly maps; and 2
CMB--only maps.  We shall refer to these sets of maps as sets (a),
(b) and (c), respectively. Set (a) is composed of five maps
corresponding to each frequency bandpasses observed by WMAP. For
each pixel in these maps, noise weighting is used to compute the
average temperature from the individual differencing assembly map
values. Set (b) consists of 'foreground--cleaned' sky maps
corresponding to each individual differencing assembly. The
assemblies are labelled $Q1$, $Q2$, $V1$, $V2$, $W1$, $W2$, $W3$ and
$W4$, where the letter corresponds to the frequency band. These maps
were used in the calculation of the angular power spectrum by the
WMAP team \cite{hsvh03}. The Galactic foreground signal was removed
using a 3--band, 5--parameter template fitting method described in
Bennett et al (2003b).  Set (c) corresponds to two CMB--only maps
constructed by the WMAP team (Bennett et al. 2003b) and Tegmark, de
Oliveira-Costa \& Hamilton (2003) (see papers for details). We shall
refer to these two maps as the ILC and TOH maps, respectively. Both
maps were assembled in a manner that minimises foreground
contamination and detector noise, leaving a pure CMB signal. The
ultimate goal of these two maps is to build an accurate image of the
last--scattering--surface that captures the detailed morphology.

For CMB analyses, it is necessary to mask out regions of strong
foreground emission. Bennett et al. 2003b provide masks for
excluding regions where the contamination level is large. The masks
are based on the $K$--band measurements, where contamination is most
severe. The severity of the mask is a compromise between eliminating
foregrounds and maximising sky area in analyses. In our analysis, we
concentrate our statistics on maps with the kp2 mask applied. This
results in 15.0\% of pixels being cut. When set (b) is combined with
the kp2 mask, we replicate the data used in the calculation of the
angular power spectrum. The Galactic sky cut is unnecessary when
looking at set (c) as the whole sky should be uncontaminated,
however, we still apply the cut for consistency and simplicity. We
also occasionally refer to results with the application of the kp0
mask which excludes 23.0\% of the sky.

\section{Implementation}
\label{sec:implementation}

In order to keep our study  as focussed as possible, our initial
multivariate test is bivariate: simply involves looking at pairs of
variates extracted from the maps. This allows us to use the same
methods on both real and harmonic space quantities as we shall
explain later in this section. The study of bivariates has the
additional benefit of reducing the computational requirements.
However, we should emphasise that all the methods outlined in
Section \ref{sec:analysis} are applicable to larger values of $p$.
For a pair of variates ($\mathv{x}_1$, $\mathv{x}_2$), we assess the
normality of the bivariate probability distribution ${\cal
P}(\mathv{x}_1,\mathv{x}_2)$. The probability distribution should
have the form outlined in Equation \ref{eqn:multigauss}. In real
space, we look at temperature pairs ($\Delta T_1$, $\Delta T_2$).
This allows us to assess the 2--point correlation function if we
extract pixel pairs across a range of scales. We randomly select 1
million temperature pairs from a given map. As there are roughly 3
million pixels, it means that there are $\sim 10^{13}$ combinations
of pairs. We found that a million pairs were sufficient to replicate
the overall distribution. We then calculate the angular separation
of each pair. The pairs are then grouped into 100 bins according to
their separation such that the bin sizes are roughly equal. Thus, we
are left with 100 bins with $n\!\sim\!10,000$. In harmonic space, we
study the distribution of the the real $\Re$ and imaginary $\Im$
parts of the spherical harmonic coefficients. This idea could be
extended to different mode functions. Spherical harmonic modes are
independent for a Gaussian random field defined on a sphere. The
effect of a cut or other mask is to correlate these models, but
only through the introduction of a linear covariance. This means
that this method could be used to study any combination of modes
even on a cut sky.

One of the other advantages of looking in harmonic space is that it
is trivial to probe differing scale lengths. We simply study the
distribution of ($\Re$, $\Im$) for a given value of $\ell$.
Consequently, we have bins with $n\!=\!\ell$. We probe scales up to
and including $\ell\!=\!600$, however, due to the small value of $n$
at low $\ell$ we ignore the results for $\ell\!<\!100$. However, the
effect of using masks on sets (a) and (b) means that the harmonic
coefficients cannot easily be calculated. Thus, we only study
harmonic space for set (c) where the whole--sky can be appropriately
studied.

We now outline how the various statistics were computed for
bivariate data. The skewness, kurtosis and D'Agostino's statistics,
applied to the marginal distributions, are trivial to calculate. As
we have already mentioned, there is an issue with distribution of
the expected kurtosis $E(b_2)$ being non--Gaussian for small $n$.
This is not an issue in real space where $n$ is very large, but, it
is worth bearing in mind in harmonic space. We could address this
issue by modelling the distribution of $E(b_2)$ using Monte Carlo
simulations. However, as this is only an issue for a small  part of
our analysis and require significantly more computation we chose not
to adopt this method. The univariate shifted--power transformation
method is more challenging to implement. The method requires a
careful choice of minimiser/maximiser. The majority of minimizers
call for the derivative of the function being minimized. The
derivative of Equation (\ref{eqn:likelihood}) is computationally
taxing. Therefore, we choose to use a 'simplex downhill' minimizer
\cite{pftv92} that requires only the actual function. The simplex
minimizer is applicable to multidimensional functions which fulfils
our need for minimizing a 2-dimensional function. Most minimizers do
not find the global minimum, instead they get 'trapped' in local
minima. We allow for this by testing our code with various initial
starting points along the function and varying the step size.
Nevertheless, there is still the possibility that we may fail to
find the true minimum, especially if the shape of the function is
complex. Indeed, evidence is found that we fail to obtain the global
minimum in some of our results (see later). However, we our not
unduly worried by this since a local minimum will result in smaller
value of $2\!\left({\cal
L_{\mathrm{max}}}(\hat{\xi},\hat{\lambda})-{\cal
L_{\mathrm{max}}}(\hat{\xi},1)\right)$ than that calculated from the
true minimum. Thus, we will fail to detect non--normality rather
than falsely claim non--normality.

The bivariate skewness and kurtosis are simple to calculate.
However, the bivariate power transformation method has the same
challenges as in the univariate case. Again, we are trying to
minimize a 2-dimensional function and so we turn to the simplex
minimizer. Obviously, by using this minimizer we inherit the same
problems as for the univariate transformation.

Finally, we look at the linearity of the temperature pairs and
($\Re$, $\Im$) pair. We add separately three non--linear terms
to our model of the data:
\begin{equation}
\begin{array}{l}
x_{2j}=A_0+A_1x_{1j}+A_2x_{1j}^2\\
x_{2j}=A_0+A_1x_{1j}+A_2x_{1j}^3\\
x_{2j}=A_0+A_1x_{1j}+A_2/x_{1j}\\
\end{array}
\end{equation}
where, as before, $x_{ij}$ are elements from the variate
$\mathv{x}_i$. Our task is now simply to assess the significance of
the coefficient $A_2$ for the three non--linear terms using Equation
(\ref{eqn:ttest}).

\section{Results}
\label{sec:results}

In this section, we present the results of our analysis on the WMAP--derived data. Throughout the section, the expectation value of a statistic $E(Q)$, if non--zero, is marked in the figures as a straight line. The dark grey areas in the plots represent the 95\% confidence regions. For example, if the distribution of $E(Q)$ is (or approximates for large $n$) a normal distribution, then the dark grey region signifies $E(Q)\pm 1.95996\sigma$. The light grey region is the 99\% confidence region. In section \ref{sec:realspace}, we display and discuss the results of applying our statistics to temperature pairs from the assembley maps, the frequency maps, the CMB--only maps and a sample Gaussian Monte Carlo (MC) map. The latter map is used as a reference guide and to reassure us that the computation of a statistic is accurate. For each map, the statistics were applied to 100 binned distributions and therefore we expect one value (for each statistic) to be outside the 99\% confidence region. In section \ref{sec:harmonicspace}, we discuss the same analysis on the spherical harmonic coefficients obtained from the CMB--only maps and the same sample Gaussian MC map. For each map, the statistics are applied to 501 distributions ($\ell=100-600$) and as such we expect 5 points beyond the 99\% confidence region.

\subsection{Real Space}
\label{sec:realspace}

\subsubsection{Marginal normality results}

\begin{figure} {
\centering{\epsfig{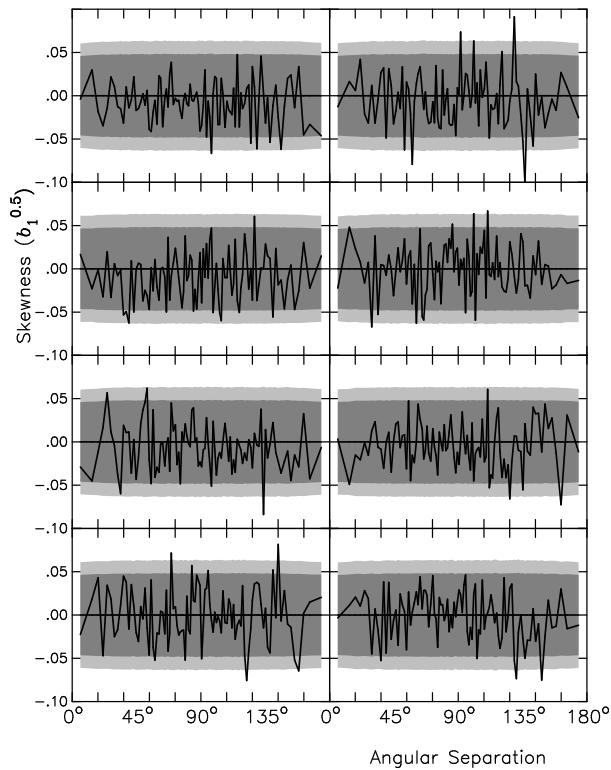}}
}
\caption{\label{fig:skew1} Skewness results from the foreground--cleaned assembly maps. ({\it Top left}) $Q$1 assembly map, ({\it top right}) $Q$2 assembly map, ({\it 2nd from top, left}) $V$1 assembly map, ({\it 2nd from top, right}) $V$2 assembly map, ({\it 3rd from top, left}) $W$1 assembly map, ({\it 3rd from top, right}) $W$2 assembly map, ({\it bottom, left}) $W$3 assembly map, and ({\it bottom, right}) $W$4 assembly map.
}
\end{figure}

\begin{figure} {
\centering{\epsfig{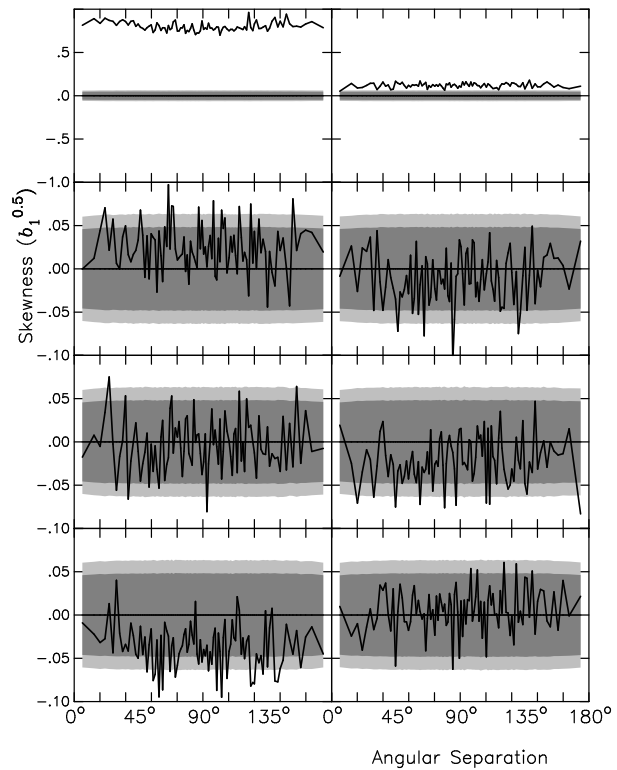}}
}
\caption{\label{fig:skew2} Skewness results from the 5 frequency maps, the 2 CMB--only maps and a Gaussian MC map.({\it Top left}) $K$ band map, ({\it top right}) $Ka$ band map, ({\it 2nd from top, left}) $Q$ band ({\it 2nd top, right})  $V$ band map, ({\it 3rd from top, left}) $W$ band map, ({\it 3rd from top, right}) ILC CMB--only map, ({\it bottom, left}) TOH CMB--only map, and ({\it bottom, right}) a Gaussian MC map.}
\end{figure}

\begin{figure} {
\centering{\epsfig{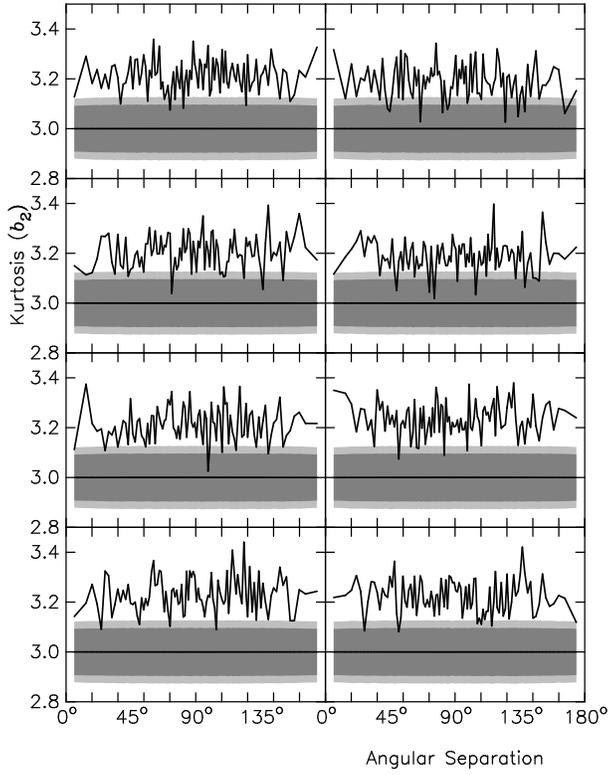}}
}
\caption{\label{fig:kurt1} Kurtosis results from the foreground--cleaned assembly maps.}
\end{figure}

\begin{figure} {
\centering{\epsfig{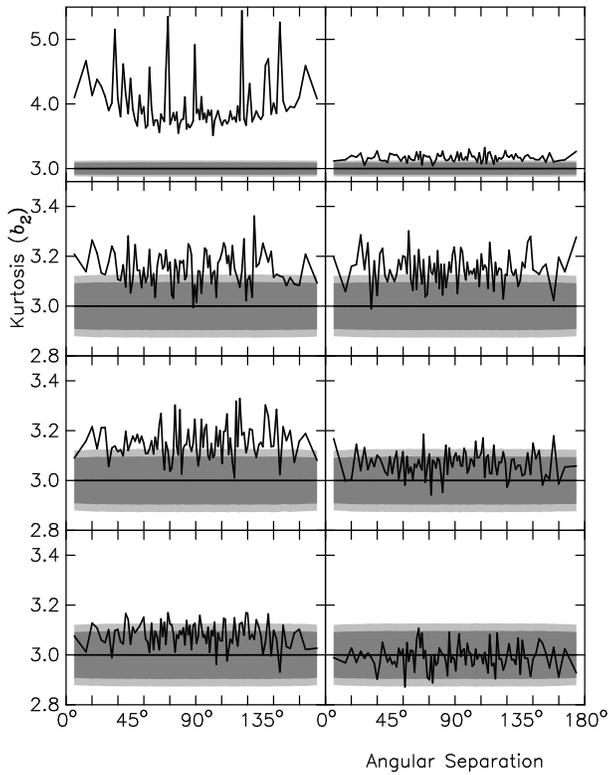}}
}
\caption{\label{fig:kurt2} Kurtosis results from the 5 frequency maps, the 2 CMB--only maps and a Gaussian MC map.}
\end{figure}

\begin{figure} {
\centering{\epsfig{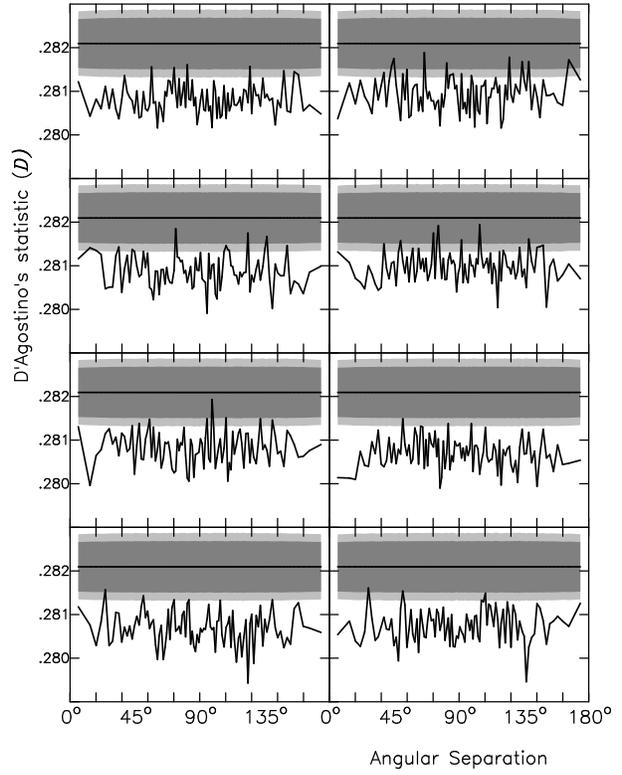}}
}
\caption{\label{fig:D1} D'Agostino's statistic results from the foreground--cleaned assembly maps.}
\end{figure}

\begin{figure} {
\centering{\epsfig{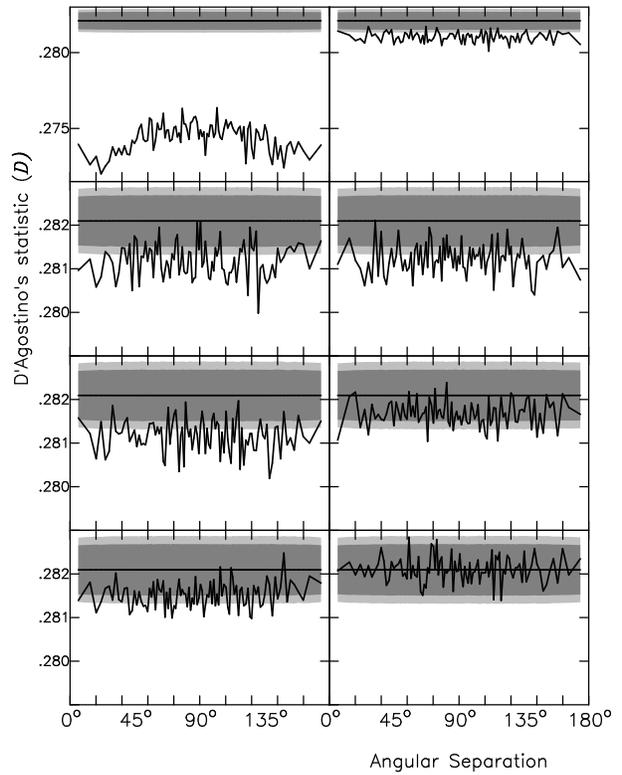}}
}
\caption{\label{fig:D2} D'Agostino's statistic results from the 5 frequency maps, the 2 CMB--only maps and a Gaussian MC map.}
\end{figure}

\begin{figure} {
\centering{\epsfig{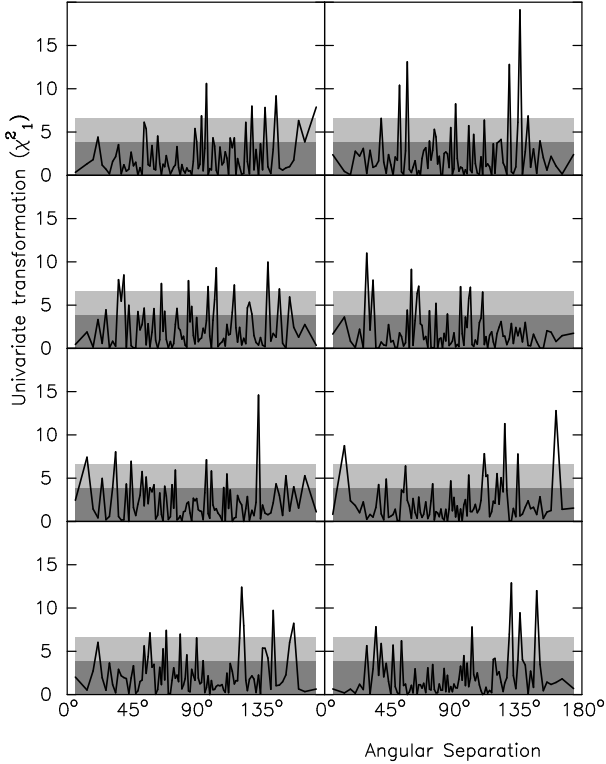}}
}
\caption{\label{fig:uni_trans1} Univariate transformation results from the foreground--cleaned assembly maps.}
\end{figure}

\begin{figure} {
\centering{\epsfig{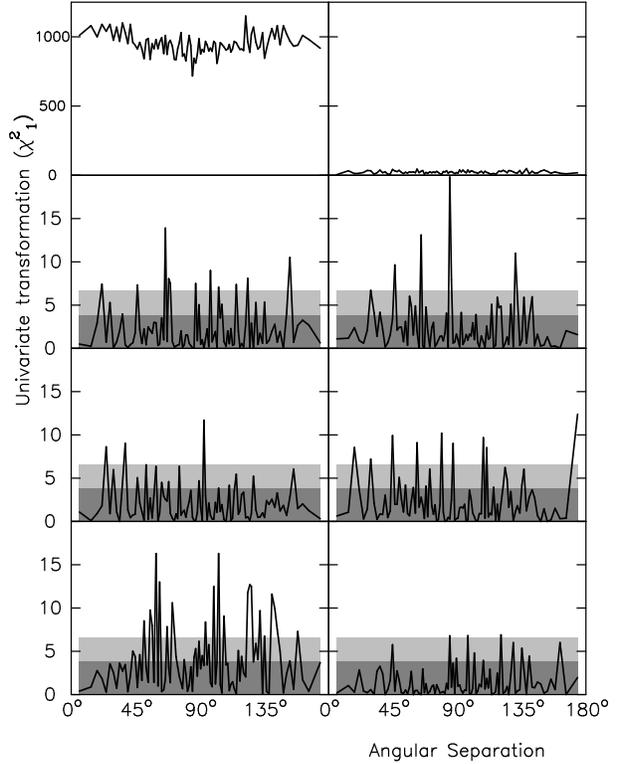}}
}
\caption{\label{fig:uni_trans2} Univariate transformation results from the 5 frequency maps, the 2 CMB--only maps and a Gaussian MC map.}
\end{figure}

The results of the application of the univariate methods on
the temperature pairs are displayed in Figures \ref{fig:skew1}-\ref{fig:uni_trans2}.
We display the results for $\Delta T_1$ to increase clarity, although unsurprising,
the corresponding plots for $\Delta T_2$ resemble those for $\Delta T_1$.

The skewness results are shown in Figures \ref{fig:skew1} and
\ref{fig:skew2}. The assembly data shows no signs of being skewed.
However, the two non--cosmological frequencies ($K$ and $Ka$) are
very skewed. Clearly, the maps are heavily contaminated with
Galactic foregrounds even outside the kp2 cut. Moreover, the
construction of the kp2 mask means that the shape of the
distribution of $K$ has an artificial cut--off at high $\Delta T$.
The skewness coefficients appear uniform across all angular
separations. Looking at the results from the cosmological
frequencies, the $Q$ band frequency appears slightly positively
skewed but the two higher frequency display no sign of
non--normality. Interestingly, when we look at the CMB--only maps,
the TOH map appears negatively skewed- there are roughly 15 points
outside the 99\% confidence region. This is the opposite sign to the
two non--cosmological bands which may suggest that the
foreground--removal process has resulted in an over-subtraction. The
ILC map also appears to be slightly negatively skewed.

The analyses of the kurtosis coefficient are shown in Figures
\ref{fig:kurt1} and \ref{fig:kurt2}. The kurtosis is higher than
expected in all 8 assembly maps and across all angular separations.
The kurtosis is even greater in the non--cosmological frequency maps
suggesting that not all the foreground information has been removed
from the assembly maps. The two CMB--only maps also hint at this but
with less certainty. Both maps have a slightly higher than expected
kurtosis especially when the two are compared to results from the
Gaussian MC map. That is to say, their distributions are too peaked,
and as with the skewness results suggests errors in foreground
modelling.

D'Agostino's test statistic is shown in Figures \ref{fig:D1} and \ref{fig:D2}.
The statistic is an omnibus statistic so it should be unsurprising that it also
pick up evidence of non--normality in all the WMAP--derived maps. Again,
the results from the assembly maps have values mainly outside the 99\% confidence region.
This shift away from the expected value is even greater for the $K$ and $Ka$ bands.
The three cosmological frequency maps produce results that appear non--normal but
with much less significance than the Galactic bands. The two CMB--only maps also
appear non--normal. At this point, we note that both the kurtosis and D'Agostino's
statistic have a greater shift away from their expected values for the Galactic
frequencies for the very large and very small angles. This feature is seen later
on in the other statistics we employ.

The results of the shifted--power transformation of the data sets are shown in Figures \ref{fig:uni_trans1} and \ref{fig:uni_trans2}. This part of our analysis was the most computationally challenging aspect, therefore, we are pleased to see that the result from the Gaussian MC map is as expected. Due to the challenge of this part of work we really look for severe departure from non--normality and keep our discussion brief. The assembly maps appear to have slightly more than expected points beyond the 99\% confidence region. Moreover, the two non--cosmological frequency maps show clear signs of non--normality. The two CMB--only maps also show signs of non--normality- the TOH map result appearing to have departed furthest from normality.

\subsubsection{Joint normality results}

\begin{figure} {
\centering{\epsfig{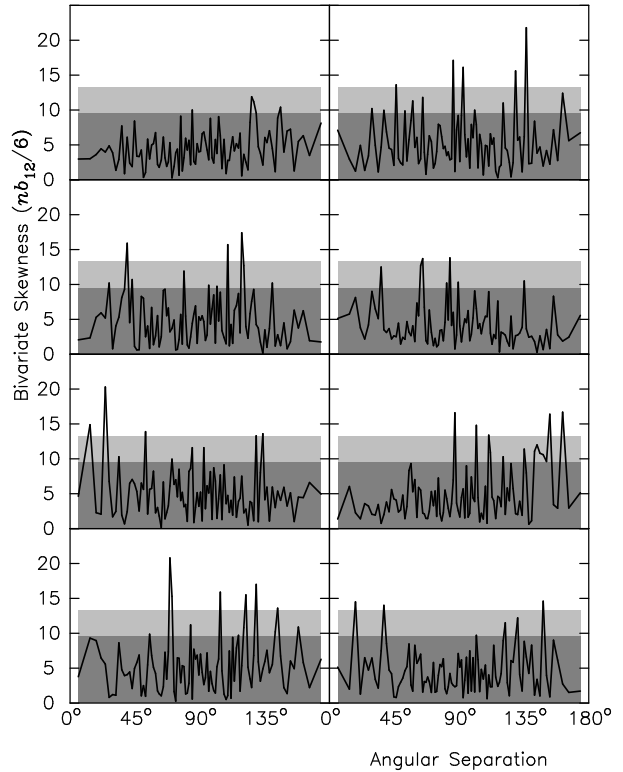}}
}
\caption{\label{fig:multi_skew1} Bivariate skewness results from the foreground--cleaned assembly maps.}
\end{figure}

\begin{figure} {
\centering{\epsfig{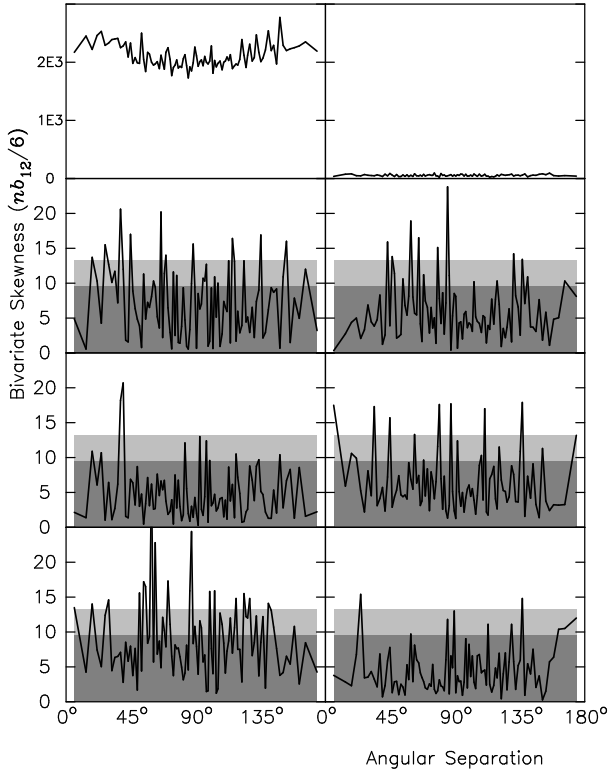}}
}
\caption{\label{fig:multi_skew2} Bivariate skewness results from the 5 frequency maps, the 2 CMB--only maps and a Gaussian MC map.}
\end{figure}

\begin{figure} {
\centering{\epsfig{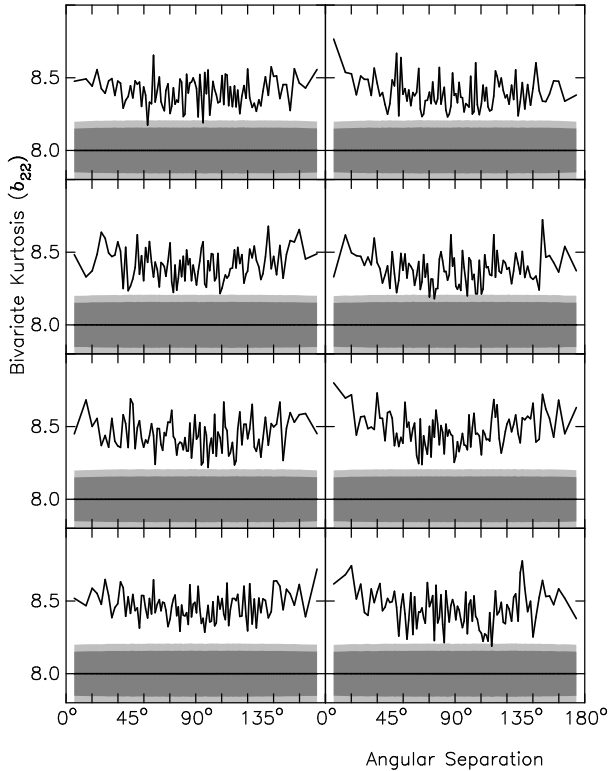}}
}
\caption{\label{fig:multi_kurtosis1} Bivariate kurtosis results from the foreground--cleaned assembly maps.}
\end{figure}

\begin{figure} {
\centering{\epsfig{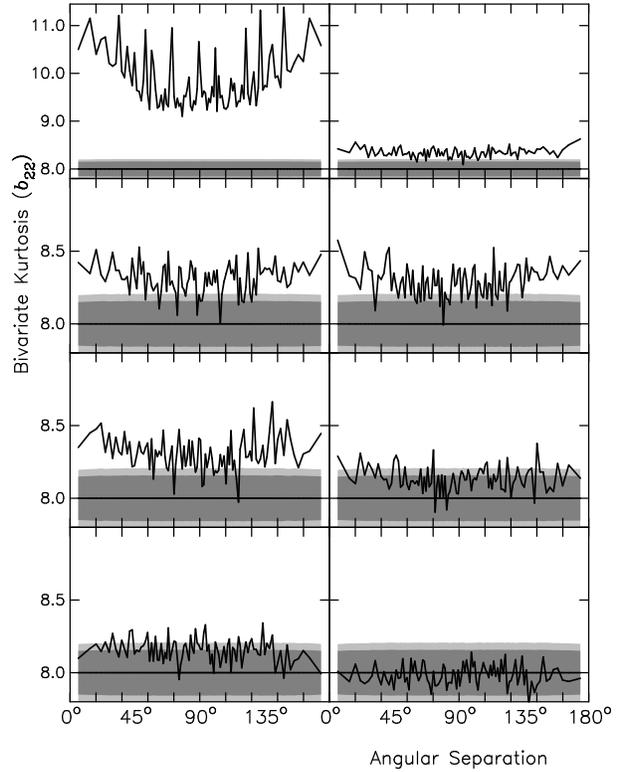}}
}
\caption{\label{fig:multi_kurtosis2} Bivariate kurtosis results from the 5 frequency maps, the 2 CMB--only maps and a Gaussian MC map.}
\end{figure}

\begin{figure} {
\centering{\epsfig{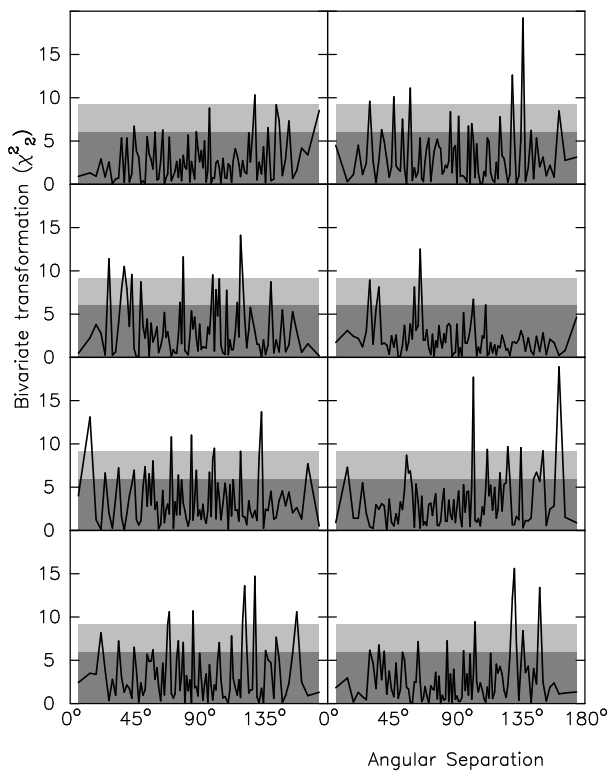}}
}
\caption{\label{fig:multi_trans1} Bivariate transformation results from the foreground--cleaned assembly maps..}
\end{figure}

\begin{figure} {
\centering{\epsfig{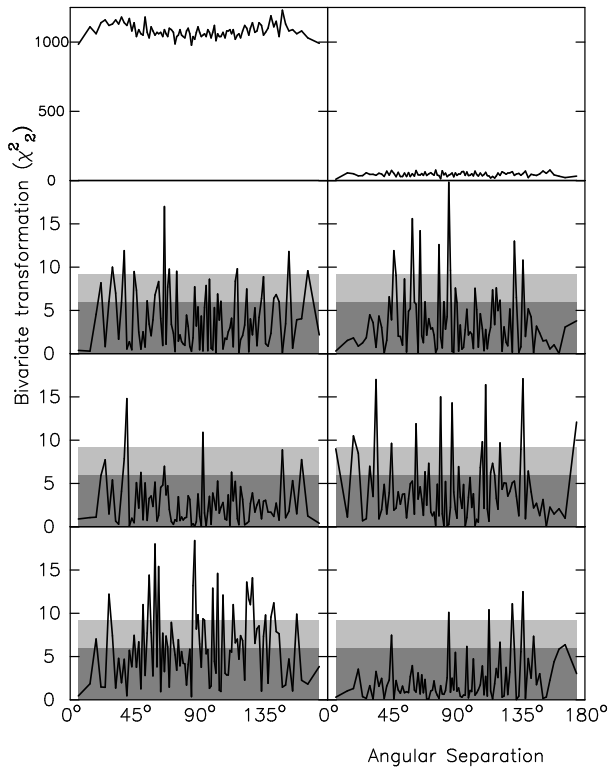}}
}
\caption{\label{fig:multi_trans2} Bivariate transformation results from the 5 frequency maps, the 2 CMB--only maps and a Gaussian MC map.}
\end{figure}

Our bivariate analysis results are shown in Figures \ref{fig:multi_skew1}-\ref{fig:multi_trans2}.
The bivariate skewness statistics ($nb_{12}/6$) calculated from the 16 maps are shown in
Figures \ref{fig:multi_skew1} and \ref{fig:multi_skew2}. The statistic appears to be
abnormal for three of the assembly maps- $Q$1, $W$2 and $W$4. Some of the other
assembly maps have two or three points outside the 99\% confidence region but
visually the results do not look too unusual.
Once again, the two non--cosmological frequencies have results that strongly
indicate non--normality. This non--normality is still evident in the $Q$ and $V$ band.
The two CMB maps also have a higher than expected number of points above the 99\% confidence region.
It would appear that the bivariate skewness results mimic their univariate counterparts.

The bivariate kurtosis results are displayed in Figures
\ref{fig:multi_kurtosis1} and \ref{fig:multi_kurtosis2}. As with the
skewness results, bivariate kurtosis seem similar to their univariate equivalent.
Nevertheless, in the case of the assembly maps, the shift away from the expected value
is even greater than for the univariate results. As before, the Galactic frequency maps
are clearly found to be non--normal. The higher value than expected value of the bivariate
kurtosis persists in the foreground--cleaned CMB maps

The bivariate power transformation is shown in Figures
\ref{fig:multi_trans1} and \ref{fig:multi_trans2}. The assembly
results do not look entirely consistent with being drawn from a
$\chi^2_2$ distribution. Five of the assembly maps produce results
with 5 or more points outside the 99\% confidence region. Saying
that, our Gaussian MC map has four points outside this region, which
makes it hard to draw definite conclusions. This is certainly not
true for the two Galactic frequency bands that are clearly
inconsistent with normality. The two CMB--only maps also appear to
have an extremely high number of points beyond the 99\% confidence
region.

\subsubsection{Linearity results}

\begin{figure} {
\centering{\epsfig{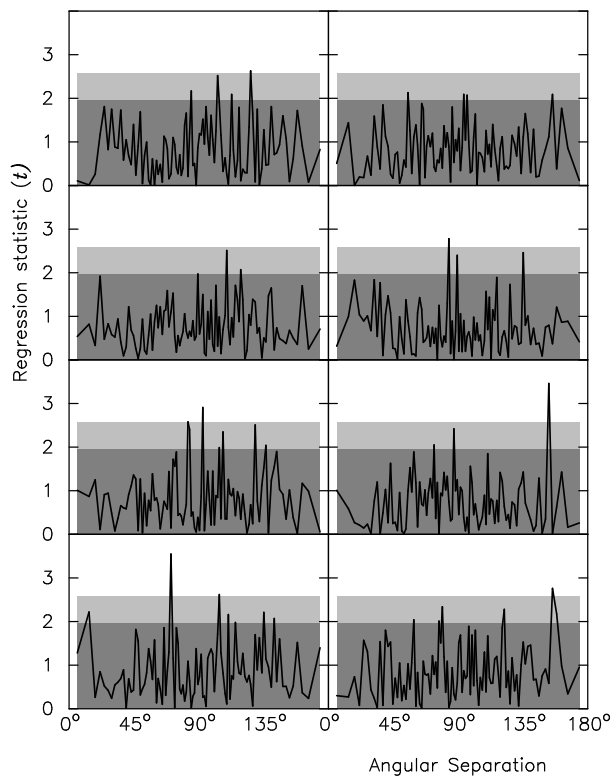}}
}
\caption{\label{fig:reg1} Linear regression results from the foreground--cleaned assembly maps.}
\end{figure}

\begin{figure} {
\centering{\epsfig{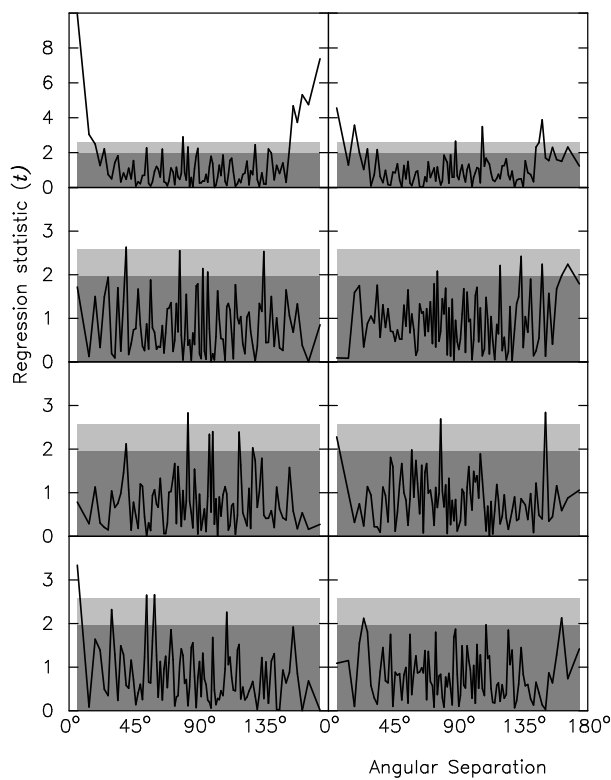}}
}
\caption{\label{fig:reg2} Linear regression results from the 5 frequency maps, the 2 CMB--only maps and a Gaussian MC map.}
\end{figure}

Lastly, in this subsection, we assess the linearity of the data. We tried adding
separately three non--linear terms to our linear model of the data (as described in
\ref{sec:implementation}). However, all three terms were found to be unnecessary
descriptors for the data from all of the maps bar the heavily contaminated $K$ band map.
This is reassuring as it tells us that the angular power spectrum supplies reliable
information about the data sets. We plot in Figures \ref{fig:reg1} and \ref{fig:reg2}
the results for addition of the $z_{j}^2$ non--linear term. Curiously, the non--linearity
seen in the $K$ band map is seen at the largest and very smallest angular separations.
As has already been noted, this trend is seen in other statistics that we employ.

\subsection{Harmonic space}
\label{sec:harmonicspace}

\subsubsection{Marginal normality results}

\begin{figure} {
\centering{\epsfig{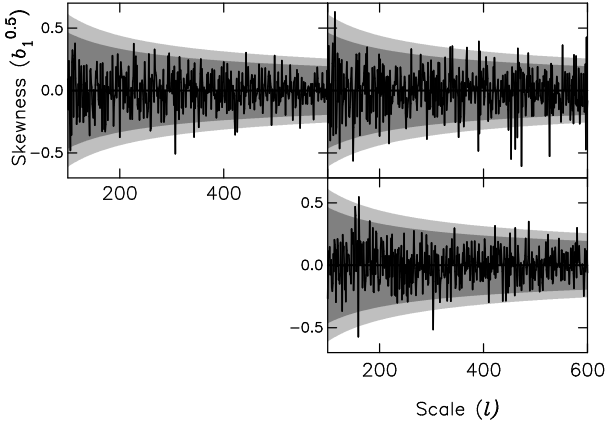}}
}
\caption{\label{fig:skewSPH} Skewness results from the $a_{\ell m}$ of the 2 CMB--only maps and a Gaussian MC map. ({\it Top left}) ILC CMB--only map, ({\it top right}) TOH CMB--only map and ({\it bottom}) a Gaussian MC map}
\end{figure}

\begin{figure} {
\centering{\epsfig{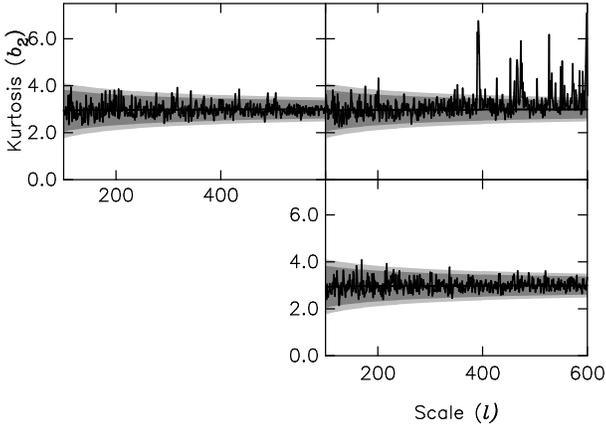}}
}
\caption{\label{fig:kurtSPH} Kurtosis results from the $a_{\ell m}$ of the 2 CMB--only maps and a Gaussian MC map.}
\end{figure}

\begin{figure} {
\centering{\epsfig{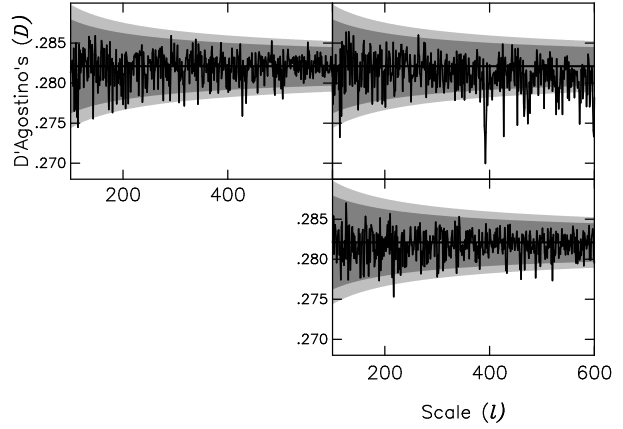}}
}
\caption{\label{fig:DSPH} D'Agostino's statistic results from the $a_{\ell m}$ of the 2 CMB--only maps and a Gaussian MC map.}
\end{figure}

\begin{figure} {
\centering{\epsfig{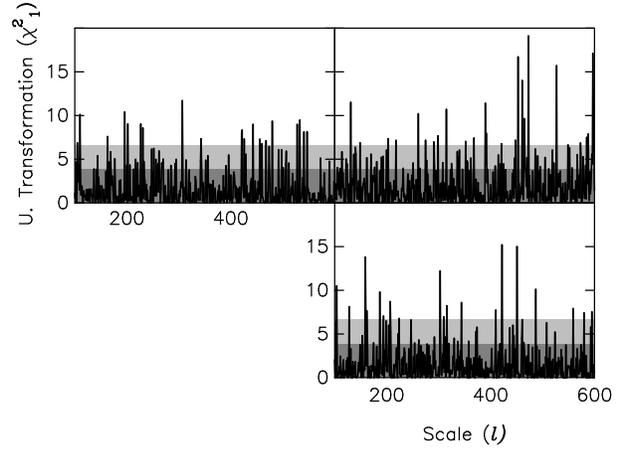}}
}
\caption{\label{fig:uni_transSPH} Univariate transformation results from the $a_{\ell m}$ of the 2 CMB--only maps and a Gaussian MC map.}
\end{figure}

The results of applying the skewness, kurtosis and D'Agostino's statistics
are shown in  Figures \ref{fig:skewSPH}, \ref{fig:kurtSPH} and \ref{fig:DSPH},
respectively. The results from the ILC map appear consistent with normality.
However, all three statistic behave unusually for the TOH map on scales
smaller than $\ell\sim 400$,. This is particularly true of the kurtosis
coefficient where the non--normality is most evident. It would be interesting
to relate this non--normality to that already seen in real space. This could
hopefully give us a better handle on the source of the non--Gaussianity
(whether Galactic or cosmological).
 The univariate transformation results are shown for completeness in
 Figure \ref{fig:uni_transSPH}. However, the method appears unreliable as
 the Gaussian MC map results are inconsistent with being drawn from a
 $\chi^2_1$ distributions. We feel this unreliability is due to the
 small values of $n$ that make the minimized function shape more complex.

\subsubsection{Joint normality results}

\begin{figure} {
\centering{\epsfig{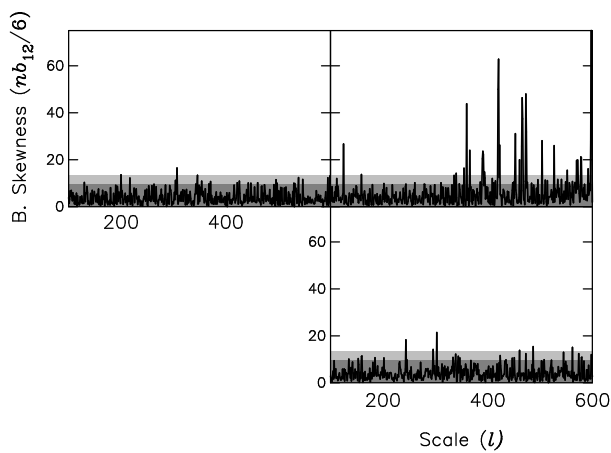}}
}
\caption{\label{fig:multi_skewSPH} Bivariate skewness results from the $a_{\ell m}$ of the 2 CMB--only maps and a Gaussian MC map.}
\end{figure}

\begin{figure} {
\centering{\epsfig{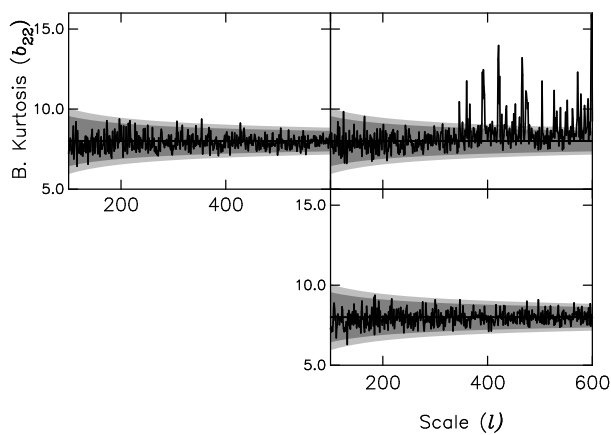}}
}
\caption{\label{fig:multi_kurtosisSPH} Bivariate kurtosis results from the $a_{\ell m}$ of the 2 CMB--only maps and a Gaussian MC map.}
\end{figure}

\begin{figure} {
\centering{\epsfig{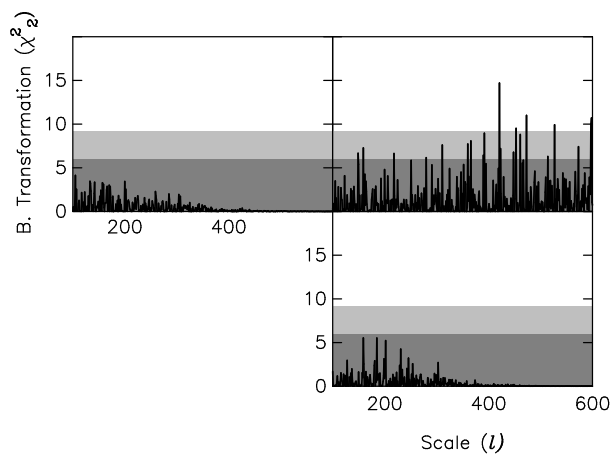}}
}
\caption{\label{fig:multi_transSPH} Bivariate transformation results from the $a_{\ell m}$ of the 2 CMB--only maps and a Gaussian MC map.}
\end{figure}

The application of our bivariate skewness and kurtosis statistics are
shown in Figures \ref{fig:multi_skewSPH} and \ref{fig:multi_kurtosisSPH}.
Once again, the ILC map behaviour corresponds to that of the Gaussian MC map.
However, the TOH map shows clear signs of non--normality. The non--normality is
displayed from larger scales than the univariate counterparts. That is to say,
the distribution of the $a_{lm}$s appears unusual at scales greater than $\ell=300$.
The results from the multivariate power transformations using harmonic space data
are shown in Figure \ref{fig:multi_transSPH}. Looking at the results from the ILC
and Gaussian MC maps, it appears that we are failing to find the true global minimum.
As discussed earlier, finding a local minimum will result in an underestimate of the
value of $\chi^2_2$. Therefore, we are not too concerned about this as failure to
find global minima since it failure does not result in false claims of non--Gaussianity.
Intriguingly, the shape of the line for the TOH map does not match those of the other
two maps. If we are failing to find the global minima then the we would expect a larger
number of points to be beyond the 99\% confidence region if we corrected this failure.
Equally, the result may reflect that the distributions extracted from the TOH maps
make finding  the global minimum easier because they are, to some extent, smoother.

\subsubsection{Linearity results}

\begin{figure} {
\centering{\epsfig{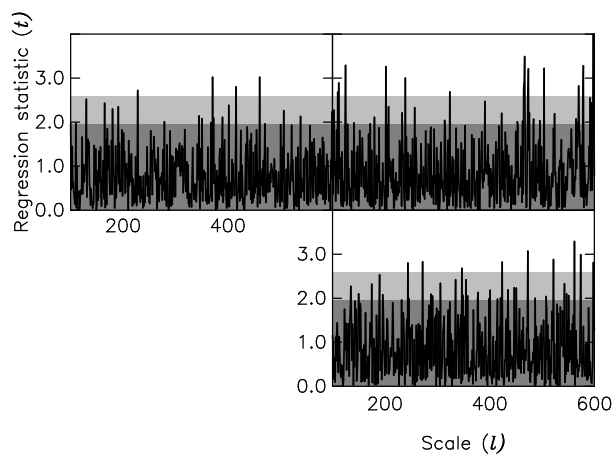}}
}
\caption{\label{fig:regSPH} Linear regression results from the $a_{\ell m}$ of the 2 CMB--only maps and a Gaussian MC map.}
\end{figure}

In our analysis of the linearity of the harmonic space variates, we do not
find any evidence of non--linearity. This was also the case when we looked
at the real space temperature pairs. To illustrate this, we display in Figure
\ref{fig:regSPH} the results from the addition of the $z_{j}^2$ non--linear term .

\subsection{Other results}

In light of the positive detections of non--normality, it is
obviously important to gain some insight into its cause. In the
Introduction, we stated that the distributions resulting from the
various sources of non--Gaussianity are poorly understood.
Nevertheless, we can reduce the influence of possible contaminants.
In particular, by applying a more severe mask to the data we should
be able to observe regions of the CMB--sky where there is less
Galactic contamination. We looked at the assembly data with the more
severe kp0 mask applied. Our techniques produce identical results to
those obtained from the kp2 mask: the skewness is consistent with
normality; the kurtosis is inconsistent ($\sim 3.2$); D'Agostino's
statistic is inconsistent (ranging from $0.280$ to $0.281$); and the
bivariate kurtosis is inconsistent ($\sim 8.5$). This suggests that
Galactic effects may not be causing the non--normality we measure.
However, we should be wary of jumping to the conclusion that the
signal is cosmological in origin. Over subtraction, residual
inhomogeneous noise, or other systematic effects may be the problem.

Another pertinent question is whether the detected non--normality is
associated with previous claims of non--Gaussianity. Eriksen et al.
(2004) find an asymmetry between the northern and southern Galactic
hemispheres, with the northern portion appearing devoid of
large--scale structure. Our techniques in real space allow us to
localize regions of the sky. We apply our statistic to the northern
and southern parts of the $W$1 assembly map. We find that the
kurtosis, D'Agostino's and bivariate kurtosis statistics show
identical signs of non--normality on both hemispheres. This suggests
that the non--normality is symmetric about the Galactic plane.
Moreover, this also rules out the non--normality being associated
with a localised 'cold--spot' as detected by wavelet techniques
\cite{mhlm05,vmbs04}.

\section{Conclusion}
\label{sec:conclusion}

In this paper, we have outlined a series of statistics that can be
used to assess the multivariate Gaussian nature of CMB data. The
extraction of cosmological parameters from this data relies on it
being jointly normal. The statistics we describe test differing
aspects of joint--normality. The first four statistics assess the
normality of marginal distributions using familiar univariate
methods. We then utilise three statistics that directly assess
joint--normality. Finally, we look for evidence of non--linearity in
the relationship between variates. We applied these tests to
bivariates extracted from maps derived from the WMAP 1st year data.
The maps consisted of 5 frequency maps, 8 'foreground--cleaned'
assembly maps and 2 CMB--only maps. The maps were assessed with the
kp2 mask applied. The bivariates extracted were temperature pairs
($\Delta T_1$, $\Delta T_2$) and the real and imaginary parts of the
spherical harmonic coefficients $a_{\ell m}$. Although, the latter
was only assessed for the two CMB--only maps.

Significant departures from normality were found in all the maps in
both real and harmonic spaces. In particular, temperature pairs
extracted from all 8 assembly maps were found to be inconsistent
with joint--normality. These maps are used to calculate the angular
power spectrum and subsequently deduce cosmological parameters.
Marginal distributions were found to have values of kurtosis and
D'Agostino's statistic outside the 99\% confidence regions.
Temperatures pairs from the same maps were also found to have values
of bivariate kurtosis outside the 99\% confidence region. These
departures were found at all angular separations. In trying to
ascertain the origin of this non--normality, we found that the
results were unaffected by the size of the Galactic cut and were
evident on either hemisphere of the CMB--sky. This last aspect rules
out the non--Gaussianity being related to previous claims of
north--south asymmetry or 'cold--spots' detected by wavelet
techniques, although residual systematics from the map-making
process remain a likely possibility for the origin of the signal.

The transformation techniques for assessing normality described in
this paper are quite challenging to implement. In future work, we hope
to improve our method such that greater confidence can be placed on
the results. One of the benefits of the transformation techniques is
that they provide a natural solution to how to modify the data such
that it is Gaussian. The positive detections of non--normality seen
in bivariate data, should make it worthwhile to assess quantities
with larger values of $p$. This will enable us to build a broader
picture of the shape of the distribution. Such knowledge should be
supplemented with advances in the understanding of distributions
from known sources of non--Gaussianity (whether cosmic or Galactic).
The techniques outlined can also be used to study other quantities
derived from the data that should conform to joint--normality. For
example, the techniques could be applied to coefficients from
wavelet or multipole vector analyses. Finally, we note that the
techniques can be incorporated into methods for subtracting sources
of non--Gaussianity. Firm requirements on the final data, such that
they satisfy these tests, will result in cleaner data--sets.

We stress that the currently-available data sets are preliminary;
the foreground subtraction for the final CMB-only maps is not
perfect, and there may well be residual systematics in the
instrument noise. Just as the data set is preliminary so is this
analysis. We look forward to further releases in order to establish
whether the non-Gaussianity that we have detected can be entirely
explained by such artifacts.

\section*{Acknowledgements}
We gratefully acknowledge the use of the HEALPix package and the Legacy
Archive for Microwave Background Data Analysis (LAMBDA). Support
for LAMBDA is provided by the NASA Office of Space Science. This work was supported by PPARC.

\end{document}